\documentclass[11pt,twoside]{article}

\usepackage{asp2006}
\usepackage{graphicx}

\begin{document}
\title{KBS 13 -- a Rare Reflection Effect sdB Binary with an M Dwarf Secondary}  
\author{B.-Q. For and H. Edelmann}   
\affil{McDonald Observatory, University of Texas at Austin, 1 University
Station, C1402, Austin, TX 78712, USA}    
\author{E.M. Green}   
\affil{Steward Observatory, University of Arizona, 933 N. Cherry Avenue,
Tucson, AZ 85721, USA}    
\author{H. Drechsel and S. Nesslinger}   
\affil{Dr.-Remeis-Sternwarte, Institute for Astronomy, University Erlangen-
Nuremberg, Sternwartstr. 7, 96049 Bamberg, Germany}
\author{G. Fontaine}   
\affil{D\'epartment de Physique, Universit\'e de Montr\'eal, C.P. 6128,
Succ. Centre-Ville, Montr\'eal, Qu\'ebec H3C 3J7, Canada}  

\begin{abstract} 
We report preliminary $VRI$ differential photometric and
spectroscopic results for KBS~13, a recently discovered non-eclipsing sdB+dM
system. Radial velocity measurements indicate an orbital
period of $0.2923 \pm 0.0004$ days with a semi-amplitude velocity of 22.82
$\pm$ 0.23 ${\rm km\,s^{-1}}$. This suggests the smallest secondary minimum
mass yet found.  We discuss the distribution of orbital periods and secondary
minimum masses for other similar systems.
\end{abstract}


\section{Introduction}   

SdB and dM binaries are fairly rare (Green et al.\ 2005), even though
it is photometrically straightforward to detect an M dwarf secondary
by its `reflection effect' for orbital periods less than a day or two.
Still, even though the peak in the orbital period histogram is close
to one day, there are about 20 to 30 times as many known post-common
envelope sdB+white dwarf binaries as there are sdB+dM binaries.  The
latter are particularly interesting because they will eventually
evolve into cataclysmic variables (CV). Understanding the pre-CV
evolution may shed additional light on, for example, the CV period
gap.

KBS~13 was selected from a list of blue objects in the fields of the
Kepler mission. D.~Sing's spectroscopic survey (priv.~communication)
of Kepler Blue Stars (KBS) identified it as an sdB. Exploratory
lightcurves for several KBS sdB candidates in November 2005 showed
a reflection effect for KBS~13.  Its 2MASS colors constrain the main
sequence companion to be no brighter than a mid-M dwarf.  Follow-up
photometry in May and September 2006 confirmed the reflection effect,
and showed that KBS~13 does not eclipse.

\section{Spectroscopic Results}

\begin{figure}
  \centering
  \includegraphics[width=0.6\linewidth]{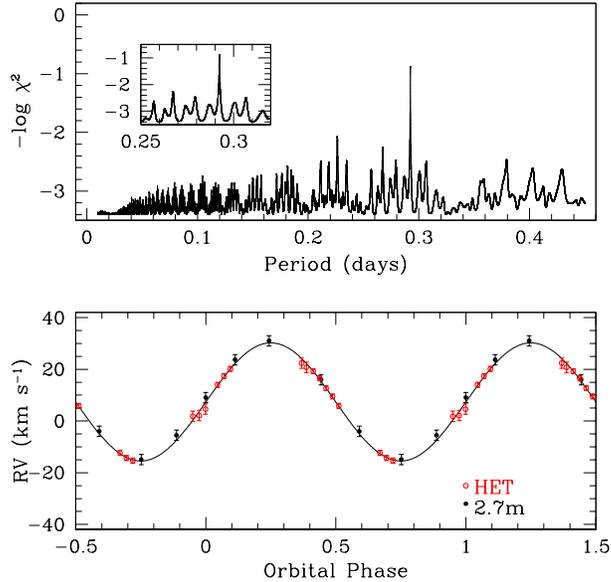}
  \caption{Top panel: power spectrum. Bottom panel: Radial velocity curve.}
  \label{fig_kbs13_rvcurve}
\end{figure}

We obtained optical high-resolution echelle spectra with the 2.7~m and
the Hobby-Eberly telescope (HET) at the McDonald Observatory.  Radial
velocities (RV's) were calculated by comparing the measured
wavelengths of all clearly identified metal lines with laboratory
values.  The 1~$\sigma$ error values are about 1~${\rm km\,s^{-1}}$.
Fig.~\ref{fig_kbs13_rvcurve} shows the $\chi^{\rm 2}$ minimization
over a range of periods, while the bottom panel shows the phased RV curve.

A zero metallicity NLTE model grid was fitted to the Balmer and Helium lines
of low-resolution spectra obtained at the Steward 2.3-m telescope
to derive $T_{\rm eff}$, $\log g$ and
$\log N({\rm He})/N({\rm H})$. Table~\ref{tab_parameters} summarizes the 
derived system parameters.

\begin{table}
\centering
\caption{Derived system parameters from spectroscopy}
\label{tab_parameters}
\smallskip
{\small
\begin{tabular}{lclc}
\tableline
\noalign{\smallskip}
Parameter & Value & Parameter & Value\\
\noalign{\smallskip}
\tableline
\noalign{\smallskip}
$T_{\rm eff}$ (K)\dotfill &  33970 $\pm$ 150 & log~$g$\dotfill & 5.87 $\pm$ 0.03 \\
log N(He)/N(H)\dotfill & $-1.60$ $\pm$ 0.04 & $R$ ($R_{\rm \odot}$)\dotfill & 0.135 $\pm$ 0.006 \\
Period (days)\dotfill & 0.2923 $\pm$ 0.0004 & $T_{\rm 0}$ (HJD)\dotfill & 2454271.9691 $\pm$ 0.0002 \\
$K$ (km\,s$^{-1}$)\dotfill & 22.82 $\pm$ 0.23 & $\gamma$ (km\,s$^{-1}$)\dotfill & 7.53 $\pm$ 0.08 \\
$M_{2,{\rm min}}$ ($M_{\odot}$)\dotfill & 0.047 $\pm$ 0.011 & $V_{\rm rot}$ (km\,s$^{-1}$)\dotfill & 23.3 $\pm$ 1.1 \\
\noalign{\smallskip}
\tableline
\end{tabular}
}
\end{table}

\section{Photometric Results}

More extensive observations in the $VRI$ band covering the orbit of KBS~13 were
carried out in 20--25 June 2007 using the Mont4K CCD on the Steward
Observatory 1.55-m telescope.  By alternating the filters, lightcurves
in two different bands were obtained each night, relative to multiple
reference stars of comparable magnitude and color to KBS~13.

\begin{figure}
  \centering
  \includegraphics[width=0.48\linewidth]{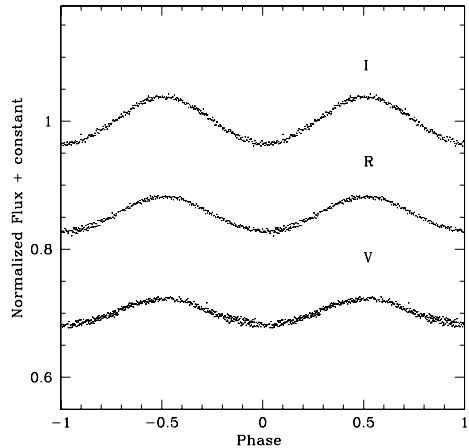}
  \caption{Phased $VRI$ lightcurves taken in June 2007 at the 1.55-m telescope.}
  \label{fig_data_fitted}
\end{figure}

We attempted to derive the orbital period from the combined 2006 and
2007 photometry.  Unfortunately, due to the large observing gap, the
Lomb-Scargle period search routine failed to give a unique solution.
Therefore, we folded the lightcurves with the orbital period and
ephemeris derived from spectroscopy.  In Fig.~\ref{fig_data_fitted},
we show the phased $VRI$ lightcurves; the amplitude variations
are $\Delta I=0.066$~mag, $\Delta R=0.048$~mag and $\Delta
V=0.035$~mag, respectively.

\section{Similar Systems}

Of the eight previously known sdB+dM binaries, PG~1017-086 (Maxted et al.\ 2002),
HS~0705+6700 (Drechsel et al.\ 2001), PG~1336-018 (Kilkenny et al.\ 1998),
HS~2231+2441 (\O stensen et al.\ 2007), HW Vir (Wood \& Saffer 1999), HS~2333+3927
(Heber et al.\ 2004), PG~1329+159 (Green et al.\ 2004) and
PG~1438-029 (Green et al.\ 2005), five have orbital periods under 3.0~h
(shorter than all but two of the more than 50 known sdB+WD binaries) and
typical secondary masses of 0.10--0.15~$M_\odot$. A sixth has a period of 4.1~h and a
minimum mass of 0.18~$M_\odot$. The remaining two sdB+dM systems have periods
of 6.0 and 8.2~h, and low velocity semi-amplitudes of 38 and 32~km~s$^{-1}$,
respectively, which imply surprisingly small minimum masses of 0.07~$M_\odot$ for
both secondaries.

\begin{figure}
  \centering
  \includegraphics[width=0.48\linewidth]{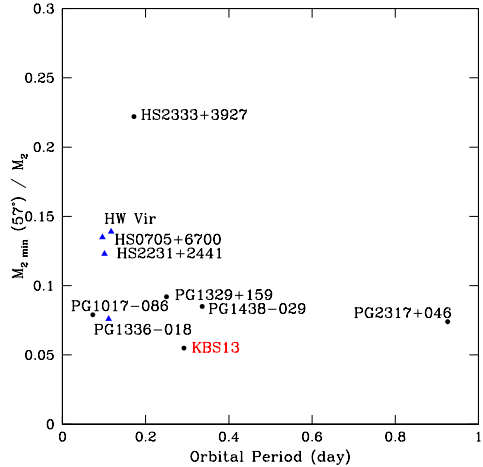}
  \caption{The distribution orbital period {\it vs} secondary minimum mass
for sdB+dM binaries. Eclipsing and non-eclipsing systems are marked by triangles and solid circles,
respectively.}  
  \label{fig_m2p}
\end{figure}

KBS~13 is only the third known sdB+dM binary with an orbital period
longer than 4.5~h.  Its secondary has the smallest minimum mass found
so far, $\sim$0.046~$M_\odot$.  Although the inclination is not yet
known, it cannot be very low for any of these systems or we would not
be able to see the reflection effects as strongly as we do.  The
distribution of orbital period {\it vs} $M_{2,{\rm min}}$ for all
known sdB+dM binaries (Fig.~\ref{fig_m2p}) shows no correlation; in
fact, the systems with the longest periods have the smallest minimum
masses.  This is unexpected, since common envelope theory indicates
that lower mass secondaries would need to spiral in much closer in
order to eject the envelope. Therefore, longer period systems
should be the ones with higher mass secondaries.

\section{Outlook}

The $VRI$ lightcurves will be analyzed simultaneously with the MORO code (Drechsel
et al.\ 1995). This program is primarily based on the Wilson-Devinney logistical
approach but the underlying model is a modified Roche model that
has taken into account the radiative interaction between the
components of hot, close binaries. Commonly, difficulties arise when the system
is degenerate in mass ratio and shows no eclipse (i.e.~low inclination). We hope
that further constraints on the free parameters will give us a unique set of
system parameters for KBS~13.

\acknowledgements 
B.-Q. For acknowledges generous travel funds from the SOC, her advisor (Chris
Sneden), the UT COX Excellence fund and the UT stellar group.



\end{document}